\documentclass[runningheads,a4paper]{llncs}

\usepackage{amssymb}
\usepackage{graphicx}
\usepackage{bbding}

\usepackage{xcolor}
\usepackage{hyperref}

\usepackage{mathtools}  
\usepackage{url}
\urldef{\mailsa}\path|{khrapov,khoperskov}@volsu.ru|
\newcommand{\keywords}[1]{\par\addvspace\baselineskip
\noindent\keywordname\enspace\ignorespaces#1}

    \def\|{\partial}
    
    \def\Oo {\displaystyle}

\begin{document}

\title{Smoothed-Particle Hydrodynamics Models: \\ Implementation Features on GPUs}

\titlerunning{Smoothed-Particle Hydrodynamics Models}

\author{Sergey Khrapov\Envelope  \ %
\and Alexander Khoperskov}
\authorrunning{S. Khrapov, A. Khoperskov}

\institute{Volgograd State University, Volgograd, Russia\\
\mailsa\\
}

\maketitle

\begin{abstract}
Parallel implementation features of self-gravitating gas dynamics modeling on multiple GPUs are considered applying the GPU-Direct technology.
The parallel algorithm for solving of the self-gravitating gas dynamics problem based on hybrid OpenMP-CUDA parallel programming model has been described in detail.
The gas-dynamic forces are calculated by the modified SPH-method (Smoothed Particle Hydrodynamics) while the N-body problem gravitational interaction is obtained by the direct method (so-called Particle-Particle algorithm).
The key factor in the SPH-method performance is creation of the neighbor lists of the particles which contribute into the gas-dynamic forces calculation.
Our implementation is based on hierarchical grid sorting method using a cascading algorithm for parallel computations of partial sums at CUDA block.
The parallelization efficiency of the algorithm for various GPUs of the Nvidia Tesla line (K20, K40, K80) is studied in the framework of galactic' gaseous halos collisions models by the SPH-method\footnote{\textcolor{blue}{Khrapov S., Khoperskov A. Smoothed-particle hydrodynamics models: implementation features on GPUs // Communications in Computer and Information Science, 2017, v.793, 266-277  https://doi.org/10.1007/978-3-319-71255-0\_21}}.

\end{abstract}

\section{Introduction}

Research of astrophysical systems applies special demands on the properties of computational fluid dynamics models.
Supersonic and hypersonic flows with the Mach number $\cal{M} \sim $ 1000, turbulence including small-scale, and magnetic fields self-consistent accounting are essential for the extragalactic astronomy, cosmology or accreting relativistic objects.

To describe star formation we have to model a multicomponent system with chemical transformations (accounting for tens or even hundreds of chemical reactions) \cite{Khoperskov-Vasiliev2014}.
These processes occur in non-stationary and non-homogeneous gravitational fields on small spatial scales \cite{Khoperskov-Moiseev2016}.
We should provide long-time calculations due to the problem rigidity taking into account fast processes and spatial small-scale inhomogeneities, when the total evolution time may exceed $ 10^7 $ integration time steps.

Let us particularly emphasize the presence of the dynamic boundaries between matter and vacuum.
The same problem appears in the case of the free water surface modeling in reservoirs at Earth's conditions \cite{khrapov-pisarev2013}.

All these and many other factors are important for modern numerical astrophysical models.

The fast calculation methods' usage for the gravitational force calculation has disadvantages due to poorly controlled errors of the acceleration in the case of approximate numerical methods (TreeCode, Fast Fourier Transform \cite{Zhang2015}, Fast Multipole Methods, etc.) that may require a large number of particles $N$.

We use the direct method (so-called Particle-Particle algorithm) for the gravitational force calculation.
Due to the low prices on the new hardware based on GPU technologies \cite{Khan2012} the direct method looks promising especially in models with a number of particles greater than 1 million.

All the features of parallel simulations are considered on the problem of gas halos collision around galaxies.
The new precise estimations of the intergalactic gas density in observations makes problem relevant.
These results are based on the observations of X-ray coronas around both elliptical and disk galaxies \cite{Zasov2017}.

Initially the Smoothed Particle Hydrodynamics (SPH) method was proposed to simulate the astrophysical gas \cite{Hwang2015,Monaghan1992}. It has shown to be efficient for various applications as well as in other fields of physics and technology.
The SPH approach occupies a significant market quota in astrophysical computational fluid dynamics and engineering applications.
It should be specially distinguished the GASOLINE code \cite{Wadsley2004}, Weakly Compressible Smoothed Particle Hydrodynamics for multi-GPUs systems \cite{Zhe2016} and gpuSPHASE for the engineering calculations \cite{Winkler2017}.
The aim of our research is the computational characteristics analysis of a parallel program for galaxies' gas self-gravitating subsystems modeling by the SPH and direct N-body methods using GPU technology.
An additional positive aspect of GPUs usage is the visualization efficiency for such processors, which is very important for multidimensional non-stationary multicomponent flows.

\section{Mathematical and Numerical Models}

\subsection{Basic Equations}

Let us consider the collision process of two galactic systems each of which includes $N_g/2$ particles gas subsystem (SPH) and $N_h/2$ component (N-body) collisionless dark halo.
The dynamics of gas particles is described by a system of differential equations:
\begin{equation}\label{Eq-dv_dt}
\frac{d \mathbf{v}_i}{d t} = - \frac{\nabla p_i }{\rho_i } + \sum_{j=1,j\neq i}^N{\mathbf{f}_{ij}}\,,
\end{equation}
\begin{equation}\label{Eq-dr_dt}
\frac{d \mathbf{r}_i }{d t} = \mathbf{v}_i \,,
\end{equation}
\begin{equation}\label{Eq-de_dt}
\frac{d e_i}{d t} = - \frac{p_i}{\rho_i} \nabla \cdot \mathbf{v}_i\,,
\end{equation}
where $N=N_g+N_h$, the radius-vector $\mathbf{r}_i(t)$ determines the position of the $i$-th particle in space, $\rho_i$, $p_i$, $e_i$, $\mathbf{v}_i$ are the mass density, gas pressure, specific internal energy, and velocity vector of the $i$-th particle, respectively.
The gravitational interaction force between $i$-th and $j$-th particles is
\begin{equation}\label{Eq-self-gravity}
\mathbf{f}_{ij} = -G  {\frac{m_j \,(\mathbf{r}_i - \mathbf{r}_j)}{|\mathbf{r}_i - \mathbf{r}_j + \delta|^3}}\,,
\end{equation}
where $G$ is the gravitational constant, $m_j$ is the mass of the particle, $\delta$ is the gravitational softening length at very short distances.

We use the quasi-isothermal model for the initial density distribution of dark matter in the halo and King model for the initial density profile in the bulge \cite{Khoperskov-Moiseev2014,Zasov2017}.
The equation of an ideal gas state is used to close the system of equations (\ref{Eq-dv_dt})--(\ref{Eq-de_dt})
\begin{equation}\label{Eq-e_p_rho}
e_i =  \frac{p_i}{(\gamma-1)\rho_i}\,,
\end{equation}
where $\gamma$ is the adiabatic index.

\subsection{The Numerical Scheme}

For the numerical integration of the hydrodynamics equations (\ref{Eq-dv_dt}) and (\ref{Eq-de_dt}) the spatial derivatives in these equations should be approximated.
In accordance with the SPH-approach \cite{Monaghan1992} for a finite number of particles $N_g$ any medium characteristic $A = \{\rho, e, \mathbf{v}\}$ and its derivatives $\nabla A$ are replaced in the flow region $\Omega$ by their smoothed values:
\begin{equation}\label{Eq-SPH-Asum}
    \begin{array}{c}
        \Oo \widehat{A}(\mathbf{r})    =   \sum_{j=1}^{N_g} {m_j \over \rho(\mathbf{r}_j)}\, A(\mathbf{r}_j)\, W(|\mathbf{r}-\mathbf{r}_j|,h) \, , \\ \\
        \Oo \nabla \widehat{A}(\mathbf{r})    =   \sum_{j=1}^{N_g} {m_j \over \rho(\mathbf{r}_j)}\, A(\mathbf{r}_j)\, \nabla W(|\mathbf{r}-\mathbf{r}_j|,h) \, , \\
    \end{array}
\end{equation}
where $W$ is the smoothing kernel function, $h$ is the smoothing length.
The following conditions are imposed on the kernel $W$:
\begin{itemize}
  \item the kernel finiteness;
  \smallskip
  \item $\Oo\int_{\Omega} W(|\mathbf{r}-\mathbf{r'}|,h)\,d\mathbf{r'} = 1$ is the normalization condition;
  \smallskip
  \item $\Oo\lim\limits_{h\rightarrow0} W(|\mathbf{r}-\mathbf{r'}|,h) = \delta(|\mathbf{r}-\mathbf{r'}|)$, где $\delta$ is Dirac delta-function.
  \smallskip
\end{itemize}


Different authors have used spline functions of different orders or Gaussian distribution for the smoothing kernel $W$ \cite{Desbrun1996,Monaghan1992,Muller2003}.
In current paper a cubic spline
\begin{equation}\label{Eq-SPH-mass-density}
\rho_i = \rho (\mathbf{r}_i) = \sum_{j=1}^{N_g} m_j \, W(|\mathbf{r}_i-\mathbf{r}_j|,h_{ij}) \,
 \end{equation}
has been used to calculate mass-density of the $i$-th particle \textit{Monaghan}:
\begin{equation}\label{Eq-SPH-Kernel}
    W(\xi,h)   =   \frac{1}{\pi h^3}
    \begin{dcases*}
        1- \frac 3 2 \, \xi^2 + \frac 3 4 \,\xi^3    ,&$0 \le    \xi   \le 1$; \\
        \frac 1 4 \,(2 - \xi)^3                 ,&$1 \le      \xi   \le 2$; \\
        0                                   ,&$\xi \ge 2$;
    \end{dcases*}
 \end{equation}
where $\xi=|\mathbf{r}_i - \mathbf{r}_j|\,/\,h$ is the relative distance from the center of the $i$-th particle, $h_{ij}=(h_i+h_j)/2$ is the effective smoothing length.
The smoothing length value for each particle depends on its mass and density as $h_i = \sigma \left( m_i / \rho_i\right)^{1/3}$,  where $\sigma$ is a constant $\sim 1.2 \div 1.3$ \cite{Lodato2011,Monaghan2005}.

If a smoothing core (\ref{Eq-SPH-Kernel}) is used to calculate the gas-dynamic forces (pressure gradient), then unphysical (numerical) particles clustering \cite{Desbrun1996} will occur in the high-pressure regions.
The latter is caused by the interaction force weakening between particles in the neighborhood of $\Oo 0 < \xi < \frac 2 3$ $\left(\Oo \lim_{\xi \rightarrow 0} \frac{\partial W}{\partial \xi}=0\right)$.
A smoothing kernel $W_p$ presenting in the following form \cite{Muller2003}:
\begin{equation}\label{Eq-SPH-Kernel-p}
    W_p(\xi,h)   =   \frac{15}{64 \pi h^3}
    \begin{dcases*}
        (2 - \xi)^3                 ,&$0 \le      \xi   \le 2$; \\
        0                                   ,&$ \xi \ge 2$;
    \end{dcases*}
 \end{equation}
eliminates clustering of particles and increases the stability of the numerical algorithm.
From equation (\ref{Eq-SPH-Kernel-p}) it follows that
$
\Oo \lim_{\xi \rightarrow 0} \frac{\partial W_p}{\partial \xi}=-\frac{45}{64 \pi h^4}
$.

Applying the SPH-approach (\ref{Eq-SPH-Asum})--(\ref{Eq-SPH-Kernel-p}) to equations (\ref{Eq-dv_dt}) and (\ref{Eq-de_dt}) we finally get:
\begin{equation}\label{Eq-dv_dt-SPH}
\frac{d \mathbf{v}_i}{d t} = - \sum_{j=1,j\neq i}^{N_g} {m_j\,\Pi_{ij}\, \nabla W_p\left(|\Delta\mathbf{r}_{ij}|,h_{ij}\right)}
+ \sum_{j=1,j\neq i}^N{\mathbf{f}_{ij}}\,,
\end{equation}
\begin{equation}\label{Eq-de_dt-SPH}
\frac{d e_i}{d t} = \frac{1}{2}\,\sum_{j=1,j\neq i}^{N_g} {m_j\,\Pi_{ij}\, \Delta\mathbf{v}_{ij} \cdot \nabla W_p\left(|\Delta\mathbf{r}_{ij}|,h_{ij}\right)}
\,,
\end{equation}
where $\Delta\mathbf{r}_{ij}= \mathbf{r}_i-\mathbf{r}_j$, $\Delta \mathbf{v}_{ij}= \mathbf{v}_i-\mathbf{v}_j$,
$\Oo \nabla W_p(|\Delta\mathbf{r}_{ij}|,h_{ij})  = \frac{\partial W_p}{\partial \xi}\,\frac{\Delta \mathbf{r}_{ij}}
{|\Delta\mathbf{r}_{ij}|}\,\frac{1}{h_{ij}}$,
$\Oo \Pi_{ij} = \frac{p_i}{\rho_i^2} + \frac{p_j}{\rho_j^2} + \nu_{ij}^a$ is the pressure force symmetric SPH-approximation ensuring Newton's third law fulfillment. The artificial viscosity $\nu_{ij}^a$ is expressed via
$$
\nu_{ij}^a = \frac{\mu_{ij}\,(\beta \, \mu_{ij} - \alpha \, c_{ij})}{\rho_{ij}}\,, \quad  \quad
\mu_{ij} =
    \begin{dcases*}
        \frac{h_{ij}\,\Delta\mathbf{r}_{ij} \cdot \Delta\mathbf{v}_{ij}}{|\Delta\mathbf{r}_{ij}|^2 + \eta \, h^2_{ij}}     ,&$\Delta\mathbf{r}_{ij} \cdot \Delta\mathbf{v}_{ij} <0$; \\
        0                                   ,&$\mathrm{else}$;
    \end{dcases*}
$$
where $\rho_{ij} = (\rho_i + \rho_j)/2$, $c_{ij} = \left(\sqrt{\gamma p_i/\rho_i} + \sqrt{\gamma p_j/\rho_j}\right)/2$ are the density and sound velocity average values for $i$-th and $j$-th interacting particles, respectively. The empirical constants $\alpha$, $\beta$ and $\eta$ determine the intensity of artificial viscosity (in our calculations their reference values are $\alpha=0.5$, $\beta=1$ and $\eta=0.1$).

A second-order accuracy method of the predictor-corrector type (the so-called leapfrog method) is used for the numerical integration of differential equations (\ref{Eq-dv_dt-SPH}), (\ref{Eq-de_dt-SPH}) and (\ref{Eq-dr_dt}).
The main steps of the leapfrog method for self-gravity SPH-models are:

\emph{(I) The velocity $\mathbf{v}_i$ and internal energy $e_i$ predictor calculations at time $t + \Delta t$:}
\begin{equation}\label{Eq-v-predictor}
\widetilde{\mathbf{v}}_i(t+\Delta t) = \mathbf{v}_i(t) + \Delta t \, \mathbf{Q}_i[\mathbf{r}(t),\mathbf{v}(t),e(t)]\,,
\end{equation}
\begin{equation}\label{Eq-e-predictor}
\widetilde{e}_i(t+\Delta t) = e_i(t) + \Delta t \, E_i[\mathbf{r}(t),\mathbf{v}(t),e(t)]\,,
\end{equation}
where $\Delta t$ is the time step, $\mathbf{Q}_i$ and $E_i$ are the right-hand side of equations (\ref{Eq-dv_dt-SPH}) and (\ref{Eq-de_dt-SPH}),  respectively.

\emph{(II) particles' spatial position calculation $\mathbf{r}_i$ at time $t+\Delta t$:}
\begin{equation}\label{Eq-r-new}
\mathbf{r}_i(t+\Delta t) = \mathbf{r}_i(t) + \frac{\Delta t}{2} \, \left[\widetilde{\mathbf{v}}_i(t+\Delta t) + \mathbf{v}(t)\right]\,.
\end{equation}
After the particles' new positions $\mathbf{r}_i (t + \Delta t)$ to be defined the density $\rho [\mathbf{r}_i (t + \Delta t)])$ is refined according to equation (\ref{Eq-SPH-mass-density}).

\emph{(III) During the corrector step the velocity, $\mathbf{v}_i$, and internal energy, $e_i$, values are recalculated at time $t+\Delta t$:}
\begin{equation}\label{Eq-v-corrector}
\mathbf{v}_i(t+\Delta t) = \frac{\mathbf{v}_i(t)+\widetilde{\mathbf{v}}_i(t+\Delta t)}{2} + \frac{\Delta t}{2} \, \mathbf{Q}_i[\mathbf{r}(t+\Delta t),\widetilde{\mathbf{v}}(t+\Delta t),\widetilde{e}(t+\Delta t)]\,,
\end{equation}
\begin{equation}\label{Eq-e-corrector}
e_i(t+\Delta t) = \frac{e_i(t)+\widetilde{e}_i(t+\Delta t)}{2} + \frac{\Delta t}{2} \, E_i[\mathbf{r}(t+\Delta t),\widetilde{\mathbf{v}}(t+\Delta t),\widetilde{e}(t+\Delta t)]\,.
\end{equation}

In general, the right-hand sides in the predictor-corrector scheme (\ref{Eq-v-predictor})--(\ref{Eq-e-corrector}) are calculated twice at the same time layer.
Since the gravitational interaction of particles (\ref{Eq-self-gravity}) depends only on the particles' positions $\mathbf{r}_i$ the calculation of the force between the particles in this approach is performed once per integration time step.
The latter allows increasing of the calculations performance about 2 times keeping the same order of accuracy for the method.

To increase the stability of the numerical method during the modeling of supersonic self-gravitating gas flows, we modified the standard SPH stability condition \cite{Monaghan1992} as follows:
\begin{equation}\label{Eq-dt-stability}
\Delta t = C_{CFL}\,\min_i \left[\frac{s_{ij}^{min}}{c_{ij}^{max}(1+1.2\alpha)+1.2\beta\mu_{ij}^{max}
         + \sqrt{s_{ij}^{min}(|\mathbf{Q}_i|+|E_i|)}}\right]\,,
\end{equation}
where $\Oo s_{ij}^{min}=\min_j |\mathbf{r}_i-\mathbf{r}_j|$, $\Oo c_{ij}^{max}=\max_j c_{ij}$, $\Oo \mu_{ij}^{max}=\max_j \mu_{ij}$.
We added the third term with the square root in the denominator of (\ref{Eq-dt-stability}) and replaced $h_{ij} \rightarrow s_{ij}^{min}$ in the numerator.
Using relation (\ref{Eq-dt-stability}), a stable calculation can be performed with larger Courant number
($0.5 \lesssim C_{CFL} < 1$) and lower artificial viscosity value.

\section{Parallel Algorithm Design}

A parallel implementation of the numerical algorithm (\ref{Eq-v-predictor})--(\ref{Eq-e-corrector}) for multiple GPUs has been performed using OpenMP-CUDA and GPU-Direct technologies.
Figure 1a presents a two-level OpenMP-CUDA parallelization scheme for $k\times$GPUs, and Figure 1b shows a data transfer scheme between GPUs based on GPU-Direct technology for NVIDIA graphics processors.
\vskip -2mm
\begin{figure}[!h]
\centering
  \includegraphics[width=0.95\hsize]{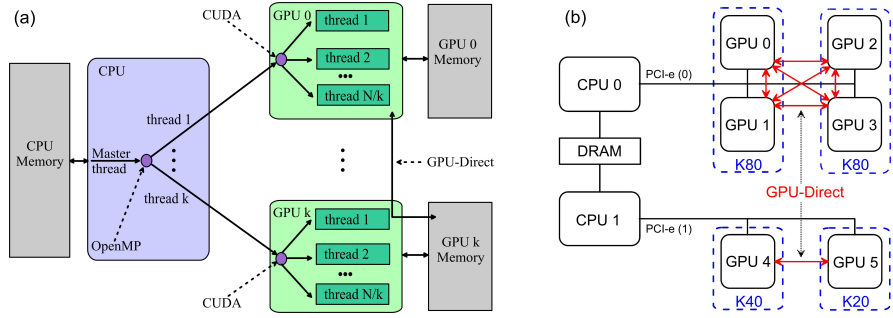}
  \caption{ (a) The two-level scheme of parallelization with OpenMP--CUDA. (b) Architecture 2$\times$CPU+6$\times$GPU.}
\label{Fig-CPUGPU}
\end{figure}
\vskip -2mm

The two-level parallelization scheme OpenMP-CUDA (Fig. 1a) is more suitable for shared memory systems type CPU + $k\times$GPU.
Using OpenMP technology to create k-threads on the CPU allows us to run CUDA kernels on $k\times$GPUs on each of which we calculate the dynamics of $N / k$ particles \cite{Dyakonova2017}.
GPU-Direct technology provides the fast data exchange between GPUs via PCI-e  bus.
This technology is only applicable to graphics processors that connect to PCI-e buses under the control of one CPU (Fig. 1b).

The numerical algorithm consists of five major Global CUDA Kernels being run from CPU on multiple GPUs using the OpenMP parallel programming model:
\begin{itemize}
  \item The Sorting Particles (SP) is a set of CUDA Kernels to determine the particles' numbering and number of particles in three-dimensional grid cells. Further this information is used to define the particles' neighbor list during the calculation of SPH sums in equations (\ref{Eq-SPH-mass-density}), (\ref{Eq-dv_dt-SPH}) and (\ref{Eq-de_dt-SPH}). The computational complexity of the kernel is $\sim O(N)$.
  \item The Density Computation (DC) is a CUDA Kernel for density calculation using (\ref{Eq-SPH-mass-density}).  It has the similar computational complexity $\sim O(N)$.
  \item The Hydrodynamics Force Computation (HFC) is a CUDA Kernel for the hydrodynamic forces calculation in (\ref{Eq-dv_dt-SPH}) and (\ref{Eq-de_dt-SPH}).  The kernel has two states \{predictor, corrector\} and its computational complexity is $\sim O(N\cdot\overline{N}_{pc})$, where $\overline{N}_{pc}$ is the average number of particles in the cells.
  \item The Gravity Force Computation (GFC) is a CUDA Kernel for the gravitational forces calculation  using (\ref{Eq-self-gravity}). The computational complexity of the kernel is $\sim O(N^2)$, because of the direct N-body method.
  \item The System Update (SU) is a CUDA Kernel for the particle characteristics updating ($\mathbf{r}_i, \mathbf{v}_i, e_i$) corresponding to equations (\ref{Eq-v-predictor})--(\ref{Eq-e-corrector}). The kernel has two states \{predictor, corrector\}, and its computational complexity is $\sim O(N)$.
\end{itemize}

\begin{figure}[!h]
\centering
\includegraphics[width=0.9\hsize]{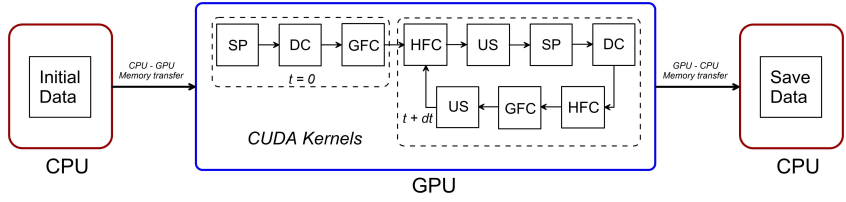}
\vskip -1mm
\caption{Flow diagram for the calculation module.}
\label{Fig-Diagrama}
\end{figure}

The Global CUDA Kernels execution sequence corresponds to the predictor-corrector scheme stages (\ref{Eq-v-predictor})--(\ref{Eq-e-corrector}). It is shown at the diagram in Figure \ref{Fig-Diagrama}.
CUDA kernels SP, DC and HFC are skipped in the case of a collisionless system.
\vskip -2mm

An important factor affecting the efficiency of the parallel implementation of the SPH method is the sorting algorithm and building a particles neighbour list.

Let us consider the algorithm parallel implementation for the particles sorting on CUDA Kernels SP in details.
The computational domain is covered by a grid $M_x \times M_y \times M_z $ with the total number of cells $M = M_x M_y M_z$.
We use the following auxiliary arrays to build particles neighbor list in the SPH method:
\begin{itemize}
  \item CellSPH$[M]$ is a vector type array of int2, the components CellSPH$[k].x$ and CellSPH$[k].y$ contain a number of particles at the $k$-cell and a number of all the particles in the $0$ to $k$ cells, respectively;
  \item indexPC$[N]$ is a vector type array of int2, the component indexPC$[i].x = k$ comprises a cell number $k$, where the $i$-th particle is located, while indexPC$[j].y = i$ links the initial number of the $i$-th particle with the sequential numeration of $j$ particles in the cells;
  \item indexCell$[M]$ is an integer type array specifying the current particle number at the corresponding $k$-th cell;
  \item maxPBC$[M/BlockSize]$ is an integer array containing the number of particles at the CUDA block, where the $BlockSize$ and $M/BlockSize$ are the CUDA block number of threads and the CUDA grid number of CUDA blocks, respectively;
  \item hmaxCell$[M]$ is a double type array comprising the smoothing length maximum value of the particle $h_i$ at the $k$-th cell.
\end{itemize}

The entire particle sorting stage contains 5 separate CUDA Kernels:
\begin{itemize}
  \item the kernelSortingSPH0$<<<$$M/BlockSize, BlockSize$$>>>$ is the NULL-initialization of sorting arrays.
  \item the kernelSortingSPH1$<<<$$N/BlockSize, BlockSize$$>>>$ includes numbers of cells where  particles are located, a number of particles and maximum value of smoothing length in cells (indexPC$[i].x$, CellSPH$[k].x$ and hmaxCell$[k]$, where
      $k = 0, ..., M-1$, $i = 0, ..., N-1$.
  \item In the kernelSortingSPH2$<<<$$M/BlockSize, BlockSize$$>>>$ the total number of particles in all the cells from $k$ to $k + BlockSize$ is defined for each CUDA block using the cascading algorithm of parallel partial sums finding. The latter is the analog of the sequential algorithm CellSPH$[k].y =$ CellSPH$[k-1].y +$ CellSPH$[k].x$.  The maxPBC$[]$ is evaluated next.
  \item Based on the total number of particles computed in the previous kernel at the CUDA block (maxPBC$[]$), in the kernelSortingSPH3$<<<$$M/BlockSize$, $BlockSize$$>>>$ the total number of particles in all cells from $0$ to $k$ is specified.
  \item In the kernelSortingSPH4$<<<$$N/BlockSize, BlockSize$$>>>$ the correspondence between the original $i$-th particle number and the sequential numbering of $j$-th particles in cells is determined (the value of indexPC$[j].x = i$ is calculated).
\end{itemize}

The fragment of the sorting algorithm code is listed below.

\medskip
\noindent
{\it The code for CUDA-core: kernelSortingSPH2}
\begin{verbatim}
__global__ void kernelSortingSPH2(int2 *CellSPH, int *maxPBC){
  __shared__ int sp[BlockSize], sp0[BlockSize];
  int ss, i, k = threadIdx.x + blockIdx.x * blockDim.x;
  sp[threadIdx.x] = CellSPH[k].x;
  sp0[threadIdx.x] = sp[threadIdx.x];      __syncthreads();
  for(i = 1; i < BlockSize; i*=2){
    if (threadIdx.x + i < BlockSize)
    sp[threadIdx.x+i] += sp0[threadIdx.x]; __syncthreads();
    sp0[threadIdx.x] = sp[threadIdx.x];    __syncthreads();}
  CellSPH[k].y = sp[threadIdx.x];
  if(threadIdx.x == 0){
    i = blockIdx.x; ss = sp[BlockSize - 1];
    while(i < gridDim.x){atomicAdd(&(maxPBC[i]), ss); i++;}}
}
\end{verbatim}

\noindent
{\it The code for CUDA-core: kernelSortingSPH3}
\begin{verbatim}
__global__ void kernelSortingSPH3(int2 *CellSPH, int *maxPBC){
  int k = threadIdx.x + blockIdx.x * blockDim.x;
  if (blockIdx.x > 0) CellSPH[k].y += maxPBC[blockIdx.x - 1];
}
\end{verbatim}

\noindent
{\it The code for CUDA-core: kernelSortingSPH4}
\begin{verbatim}
__global__ void kernelSortingSPH4(int2 *indexPC, int2 *CellSPH,
                                  int *indexCell){
  int i = threadIdx.x + blockIdx.x * blockDim.x;
  int k = indexPC[i].x, j = (k>0) ? CellSPH[k - 1].y : 0;
  int ibk = atomicAdd(&indexCell[k], 1), indexPC[j + ibk].y = i;
}
\end{verbatim}

In the CUDA kernels, DC and HFC, arrays indexPC[], CellSPH[] and $\textrm{hmaxCell}[]$ are used to find the particles neighbor list upon SPH sums calculation.

\section{The Principal Results and Discussions}

We have studied the parallelization efficiency of our algorithm solving the relevant  problem of galactic gaseous halos collisions modeling.
The calculations have been carried out on GPU Nvidia Tesla processors: K20 (1GPU), K40 (1GPU), K80 (2GPU).

A different amount of gas $N_g = N / 2$ and collisionless $N_h = N / 2$ particles has been used in the calculations. The total number of particles $N = N_g + N_h$ has been set in the range from $2^{18}$ to $2^{23}$.
\begin{figure}[!h]
\centering
\includegraphics[width=0.99\hsize]{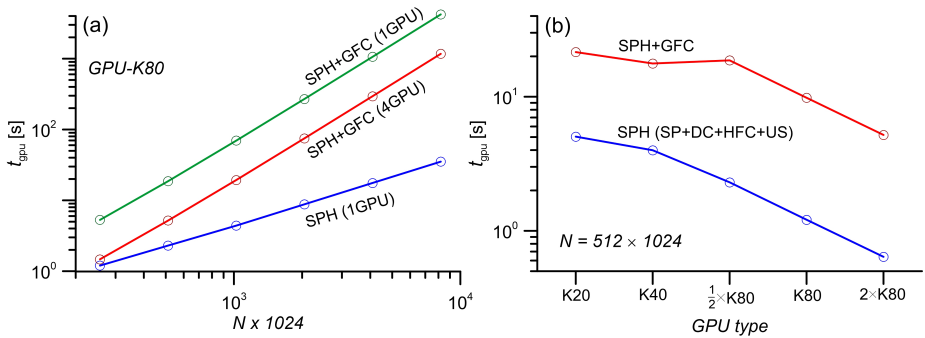}
\vskip -0mm
\caption{The execution time of CUDA kernels SPH (SP, DP, HFC, US) and GFC on GPUs. The dependence of $t_{gpu}$ on (a) the number of the particles $N$; (b) the GPU type.}
\label{Fig-Time-GPU}
\end{figure}

Figure \ref{Fig-Time-GPU} represents the computation time of the hydrodynamic and gravitational interaction of the particles for different amount of $ N $ and GPUs types.
For CUDA kernel GFC the calculation time dependence on the number of particles is almost quadratic which corresponds to the Particle-Particle algorithm complexity $O(N^2)$.
The SPH calculation time has almost a linear dependence on the number of particles, which also corresponds to the kernel HFC CUDA algorithm complexity $\sim O(N\cdot \overline{N}_{pc})$ ($\overline{N}_{pc}\simeq \textrm{const}$, since $h\sim N^{-1/3}$).
The parallelization efficiency of the algorithm on two and four GPUs is 95\% and 90\%, respectively.
\begin{table}[!h]
 \begin{center}
       \caption{The execution time of CUDA kernels SPH and GFC on GPUs.}
       \begin{tabular}{|c|c|c|c|c|c|c|}

        \hline
        \  &
        \multicolumn{2}{|c|}{$K20$ (1GPU)} &
        \multicolumn{2}{|c|}{$K40$ (1GPU)} &
        \multicolumn{2}{|c|}{$\Oo\frac{1}{2}\times K80$ (1GPU)} \\
        \hline
        $N\times 1024$  &
        $t_{SPH}$       &
        $t_{GFC}$       &
        $t_{SPH}$       &
        $t_{GFC}$       &
        $t_{SPH}$       &
        $t_{GFC}$       \\
        \hline

        256  & 2.54 & 4.13 & 2.02 & 3.51 & 1.20 & 4.10  \\
        \hline

        512  & 5.04 & 16.50 & 3.99 & 13.72 & 2.30 & 16.37  \\
        \hline

        \parbox[t]{0.13\hsize}{\centering 1024}  &
        \parbox[t]{0.13\hsize}{\centering 9.93}  &
        \parbox[t]{0.13\hsize}{\centering 65.90}  &
        \parbox[t]{0.13\hsize}{\centering 7.87}  &
        \parbox[t]{0.13\hsize}{\centering 53.62}  &
        \parbox[t]{0.13\hsize}{\centering 4.40}  &
        \parbox[t]{0.13\hsize}{\centering 65.32}  \\
        \hline

         \end{tabular}
        \label{table1}
    \end{center}
\end{table}

Table 1 shows some numerical values of the execution time of CUDA kernels SPH and GFC on different GPUs as a function of the number of particles $N$.
Figure 3b and Table 1 show that the runtime of CUDA kernels SPH (SP + DC + HFC + US) on one K80 GPU is 1.7 times less than for the K40 GPU, but the CUDA kernel GFC runs 1.2 times faster on the GPU K40.
The SPH algorithm uses only global GPU memory, and the calculation of forces between the $i$-th and $j$-th particles in the CUDA kernel GFC is organized using shared memory of the GPU.
Therefore, the different speed of CUDA kernels SPH and GFC execution on GPUs may be due to more efficient access to global memory on the K80.
\begin{figure}[!h]
\centering
\includegraphics[width=0.45\hsize]{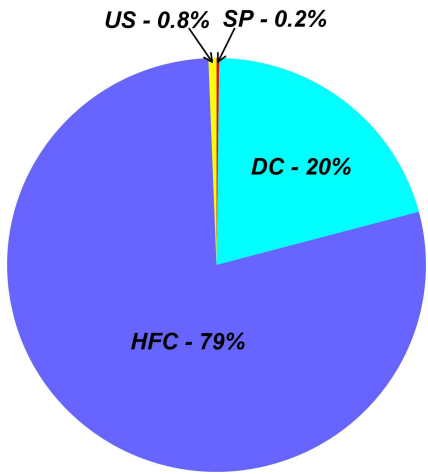}
\caption{The contributions of the different stages of the SPH numerical scheme at given time step.}
\label{Fig-Time-SPH}
\end{figure}
%

Figure \ref{Fig-Time-SPH} demonstrates that the SP sorting time is borrowed only $0.2\%$ of the total SPH simulation time.
The sorting algorithm parallel implementation on GPUs proposed in current article requires less computational and memory resources in comparison with tree-based and hash-tables algorithms \cite{Mokos2015}.
Note that the integration time step decreases ($\Delta t \sim h \sim N^{- 1/3}$) with an increase in the particles number in accordance with the stability condition (\ref{Eq-dt-stability}).
Therefore, the total time for modeling the self-consistent dynamics of particles of the gas and collisionless subsystems has a stronger dependence on $N$ than the one shown in Figure~\ref{Fig-Time-GPU}a: $t_{all}=t_{SPH}+t_{GFC}=O(N^{4/3})+O(N^{7/3})=O(N^{7/3})$.
\begin{figure}[!h]
\centering
\includegraphics[width=0.95\hsize]{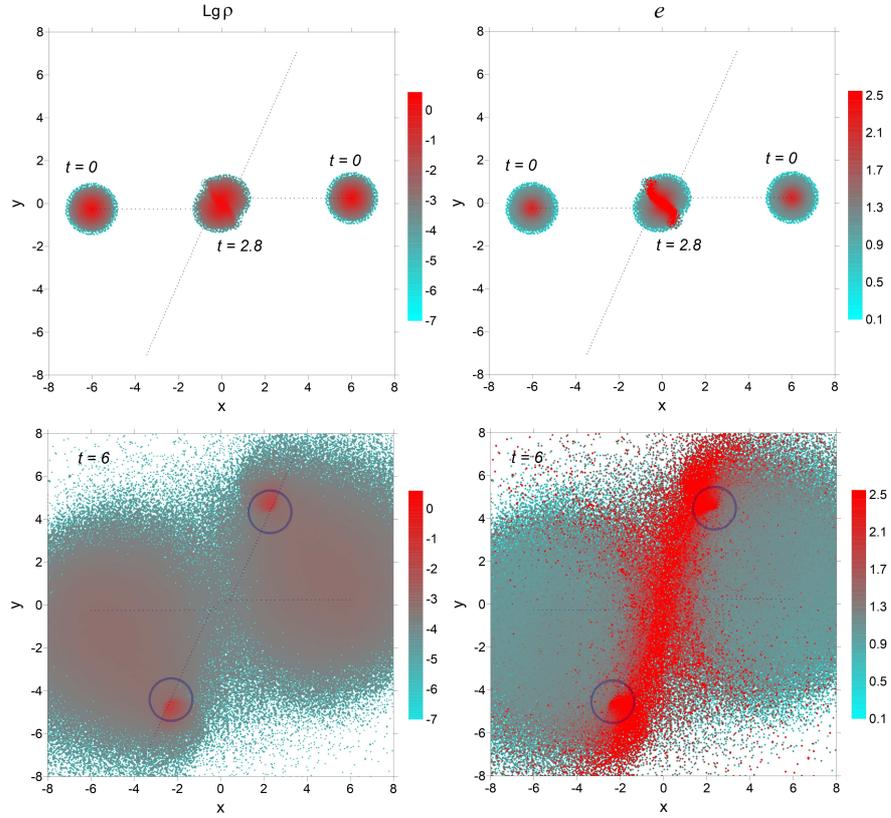}
\caption{The distribution of density (left) and internal energy (right) at different times. The dotted line and circles indicate the trajectory of the dark halo.}
\label{Fig-dynamic-SPH}
\end{figure}
%

The results of our simulation are presented in Figure~\ref{Fig-dynamic-SPH}.
In the process of collision of galaxies, there is a mixing of matter of two galactic systems.
An important factor in the interaction of galaxies is the formation of nonstationary shock waves in the collision of gas halos, leading to a substantial heating of the gas in the halo.
After the passage of the gas halos, some of their matter is emitted into the surrounding space with the formation of clouds with a nonzero angular momentum.

\subsubsection*{Acknowledgments.}
 The first author is thankful to the RFBR (grants 16-07-01037, 15-02-06204 and 16-02-00649).
 The second author has been supported by the Ministry of Education and Science of the Russian Federation (government task No.2.852.2017/4.6).

\end{document}